%% file: main.tex
\documentclass{article}

\usepackage{arxiv}

\usepackage{wrapfig}

\usepackage[utf8]{inputenc} 
\usepackage[T1]{fontenc}    
\usepackage{hyperref}       
\usepackage{url}            
\usepackage{booktabs}       
\usepackage{amsfonts}       
\usepackage{nicefrac}       
\usepackage{microtype}      
\usepackage{lipsum}

\usepackage{chngcntr}\usepackage{graphicx}
  
\usepackage{dcolumn}
\usepackage{stackengine}
\usepackage[section]{placeins}
\usepackage[dvipsnames*,svgnames]{xcolor}
\usepackage{verbatim}
\usepackage[position=bottom,caption=false,captionskip=3pt,font=normalsize,subrefformat=simple,labelformat=simple]{subfig}

\usepackage{listings}
\usepackage{mathtools}
\usepackage{todonotes}
\usepackage{amsmath}
\usepackage{chngcntr}
\usepackage{bm}
\usepackage{algorithm}
\usepackage{algpseudocode}

\usepackage{wrapfig}

\usepackage[super]{nth}
\usepackage[euler]{textgreek}

\usepackage{caption}

\hypersetup{
    colorlinks=true,
    linkcolor={blue},
    filecolor={blue},      
    urlcolor={blue},
    citecolor={blue}
}

\hypersetup{colorlinks,linkcolor={blue},citecolor={blue},urlcolor={black}}

\definecolor{GRgreen}{rgb}{0.2, 0.55, 0}
\definecolor{AGpurple}{rgb}{0.5, 0.1, 0.7}
\newcommand{\AG}[1][]{\textcolor{AGpurple}}

\usepackage{authblk}

\title{Enhanced Thermoelectric ZT in the Tails of the Fermi Distribution via Electron Filtering by Nanoinclusions — Model Electron Transport in Nanocomposites}

\author[1,$\dagger$]{S. Aria Hosseini}
\author[1,$\dagger$]{Devin Coleman}
\author[2]{Sabah Bux}
\author[1,*]{P. Alex Greaney}
\author[1,*]{Lorenzo Mangolini}

\affil[1]{Department of Mechanical Engineering, University of California, Riverside, Riverside, CA 92521, USA}
\affil[2]{Thermal Energy Conversion Research and Advancement Group, Jet Propulsion Laboratory/ California Institute of Technology, Pasadena, CA 91109, USA}
\affil[$\dagger$]{These authors contributed equally to this work}
\affil[*]{Correspondence: PAG (Theory): greaney@ucr.edu; LM (Experiment): lmangolini@engr.ucr.edu}

\begin{document}
\maketitle

\begin{abstract}
\input{Sections/abstract}
\end{abstract}


\input{Sections/introduction}
\input{Sections/material_synthesis}
\input{Sections/thermoelectric_transport_properties}
\input{Sections/model_transport_coefficients_in_bulk_thermoelectrics}
\input{Sections/model_transport_in_nanocomposite_thermoelectrics}
\input{Sections/model_validation_against_experiment}
\input{Sections/maximum_theoritical_power_factor_enhancement}
\input{Sections/model_prediction_for_electron_thermal_conductivity}
\input{Sections/conclusion}

\section{ACKNOWLEDGEMENTS}
Authors DC and LM were supported by the National Science Foundation under grant 1351386. Part of this work was performed at the California Institute of Technology/Jet Propulsion Laboratory under contract with the National Aeronautics and Space Administration and was support-ed by the NASA Science Missions Directorate’s Radioisotope Power Systems Program.
\section{DATA AVAILABILITY}
The raw data from these measurements and calculations is available upon request from the corresponding authors.
\section{CONFLICT OF INTEREST}
The authors declare no conflict of interest.

\bibliographystyle{unsrt}  
\bibliography{references} 


\end{document}

%% file: Sections/abstract.tex
Silicon carbide nanoparticles with diameters around 8 nm and with narrow size distribution have been finely mixed with doped silicon nanopowders and sintered into bulk samples to investigate the influence of nanoinclusions on electrical and thermal transport properties.  We have compared the thermoelectric properties of samples ranging from 0–5\% volume fraction of silicon carbide. The silicon carbide nanoinclusions lead to a significant improvement in the thermoelectric figure of merit, ZT, largely due to an enhancement of the Seebeck coefficient. A semiclassical Boltzmann transport equation is used to model the electrical transport properties of the Seebeck coefficient and electrical conductivity. The theoretical analysis confirms that the enhancements in the thermoelectric properties are consistent with the energy selective scattering of electrons induced by the offset between the silicon Fermi level and the carbide conduction band edge. This study proves that careful engineering of the energy-dependent electron scattering rate can provide a route towards relaxing long-standing constraints in the design of thermoelectric materials.

%% file: Sections/introduction.tex
\section{INTRODUCTION}

The imperative for reducing global use of energy from fossil fuels is incontrovertible, and humanity is faced with the difficult task of sharply reducing its consumption of hydrocarbon deposits while the energy demand continues to increase as the world becomes more industrialized. The societal and economic hurdles to reducing energy use are rendered less painful by using energy more efficiently. Towards this end, realizing good thermoelectric (TE) performance in bulk materials that are abundant, inexpensive, and environmentally benign is a holy grail of renewable energy technologies and has the potential to transform our use, and reuse, of energy~\cite{hosseini2021mitigating, tang2010holey}.

Designing materials for efficient thermoelectric energy conversion is a far-from-trivial task that requires careful optimization of several design parameters, such as doping level, charge carrier concentration, and thermal conductivity. Thermal to electric power conversion efficiency is described by the dimensionless figure of merit, $ZT=(\sigma S^2)/(\kappa_e+\kappa_l ) T$, where $\kappa_e$ is the electrical contribution to the thermal conductivity, $\kappa_l$ is lattice thermal conductivity, $\sigma$ is the electrical conductivity and $S$ is the Seebeck coefficient (thermopower) and $T$ is the temperature~\cite{gayner2020energy}. For a given material, it is challenging to decouple the transport terms independently. For instance, doping can increase electrical conductivity, decrease thermal conductivity via electron-impurity scattering, but decrease the Seebeck coefficient. Much of the research on thermoelectrics has focused on (a) the search for materials with inherently low thermal conductivity, such as skutterudites and chalcogenides~\cite{ma2013composite}, and/or (b) the control of nanoscale features to hinder thermal transport by phonons without affecting electronic transport properties~\cite{hosseini2022universal, hosseini2021enhanced, de2020heat, de2019large, bux2009nanostructured}. While promising, these approaches rely either on materials that can be rare and expensive, limiting their potential for large-scale terrestrial applications, or on the control of nanoscale features such as diameter and length of nanowires, which also poses synthetic difficulties for large scale implementation. Here we present our most recent results in the use of nanoscale additives for the improvement of thermoelectric performance in common semiconductor materials such as silicon along with a theoretical model to elucidate them. The model strongly indicates that at least part of the improvement in thermoelectric performance observed experimentally originates from the mechanism of electron energy filtering -- an approach to enhancing a material’s thermopower by selectively scattering low energy electrons to recuperate the damage to the electrical properties~\cite{hosseini2021mitigating, gayner2020energy} -- and provides a theoretical framework for guiding the further experimental synthesis of these materials. Bulk silicon is not an efficient thermoelectric material due to its high thermal conductivity~\cite{harter2019prediction, shanks1963thermal}; however, it provides an excellent platform for studying the role of design parameters on transport properties, since its bulk properties are extremely well characterized. This study is partially motivated by our recent findings suggesting that oxide inclusion, spontaneously nucleated during the sintering of silicon nanoparticles, can be effective at improving thermoelectric power conversion~\cite{coleman2018thermoelectric}. While the mechanism is attractive, this synthesis route is problematic since the thermodynamically driven nucleation of oxide inclusions is difficult to control, meaning that inclusions' size and density are not easily and independently tunable. Here we use silicon carbide nanoparticles as an additive that is mechanically mixed via ball milling with silicon feedstock powder. The addition of even a minor quantity of silicon carbide nanoparticles (5\% by volume) increases the overall performance significantly. Careful transport measurements, coupled with detailed modeling of the electronic transport properties, unequivocally confirm that energy-selective electron scattering is responsible for the performance enhancement. The energy-selective scattering increases the Seebeck coefficient and the overall power factor (PF), defined as $\sigma S^2$. In this SiC/Si system, the improvements in power factor are relatively modest, however, using the same modeling approach we show that there is scope to achieve significant enhancements in power factor through electron energy children. Our study suggests that carefully designed nanoinclusions can enable one to increase doping concentration without the usual decrease in the Seebeck coefficient. As such, the approach overcomes long-standing intrinsic constraints that have limited the power conversion efficiency of thermoelectric materials.

%% file: Sections/material_synthesis.tex
\section{MATERIALS SYNTHESIS}

\begin{figure*}[t]
\begin{center}
\includegraphics[width=0.75\textwidth]{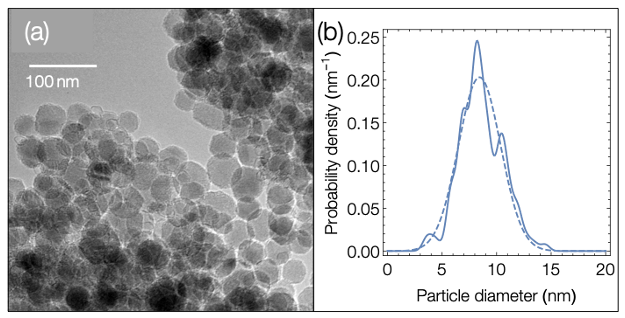}
\end{center}
\caption{Characterization of SiC nanoparticles. Panel (a) shows a TEM image of the as synthesized SiC nanoparticles with their size distribution shown in panel (b). In (b) the solid line is the measured sized distribution, and the dashed line is the best Gaussian fit to the distribution.}
\label{fig:tem}
\end{figure*}

Freestanding silicon carbide nanoparticles with a narrow distribution of sizes have been produced using the reactor described in detail in~\cite{lopez2016situ}. This synthetic approach leverages the well-established capability of non-thermal plasmas to (a) nucleate and grow small particles thanks to their electrostatic stabilization~\cite{kortshagen2016nonthermal} and (b) produce high-quality nanocrystals even for high melting point materials because of the exothermic reaction occurring at the surface of these particles while in the plasma~\cite{etherington2019persistent}. The reactor consists of two non-thermal plasmas placed in series. The first step is composed of a 2.54 cm outer diameter quartz tube, through which 320 sccm of the argon-silane mixture has flowed at 2 Torr. The silane concentration was 1.37\% by volume. Silane is fully converted into silicon nanoparticles in the first plasma stage. The second stage of the reactor is also composed of a 2.54 cm diameter quartz tube, separated from the first reactor by a 4 mm diameter orifice. The pressure in the downstream reactor is held constant at 1 Torr. 3.2 sccm of methane is flowed through an inlet immediately downstream of the orifice between the first and the second plasma. Each discharge is sustained using copper ring electrodes by a constant radio frequency (13.56 MHz) power of 100 W. The particles are collected on a stainless-steel mesh downstream of the reactor and transferred air-free to an argon-filled glovebox. The silicon carbide particles produced are roughly spherical in shape with an average diameter of 8 nm. Representative transmission electron microscopy (TEM) for these particles is shown in Fig.~\ref{fig:tem}(a). The particle size distribution obtained from the analysis of several TEM images is shown in Fig.~\ref{fig:tem}(b).

After collection, the SiC nanocrystals are mechanically mixed with silicon nanopowder at various ratios and sintered via hot pressing. The silicon nanopowder is prepared using a well-established approach of sealing silicon ingot fragments, red phosphorus, and gallium phosphide in a 50 ml tungsten carbide vial with tungsten carbide balls of 10 mm diameter, followed by Spex milling for 24 hours~\cite{bux2009nanostructured}. Doping concentrations are 2\% red phosphorus and 0.5\% GaP. The SiC nanocrystals are mixed with the silicon powder at 0\%, 1\%, 5\%, and 10\% volume fractions. After mechanical mixing, the powder is loaded into a 12.7 mm inner diameter graphite dies with boron nitride dry lubricant and sintered in a hydraulic hot press at 1160~\textdegree C and 120 MPa with a linear heating rate of $\sim$ 20~\textdegree C/min to 1160~\textdegree C and a hold time of 30 minutes. After sintering, the pressure is released, and the system is passively cooled to room temperature over a period of a few hours. The consolidated pucks are then cut and polished for characterization. Bulk samples with $>99\%$ theoretical density are achieved for the 0\%, 1\%, and 5\% samples. We found that the 10\% sample exhibits significant porosity with only $\sim$ 90\% of the theoretical density. This suggests that the addition of the silicon carbide nanoinclusions negatively affects the densification kinetics. While it is in principle possible to counteract this effect by tuning the sintering temperature and press profile, for the remainder of this manuscript we focus on the samples sintered at 1160~\textdegree C and leave a more focused study on the sintering of these composites for a future contribution.

X-ray Diffraction (XRD) is performed by a PANalytical EMPYREAN diffractometer with a CuK\textsubscript{\textalpha} source. The diffraction spectra of the consolidated pucks confirm the presence of {\textbeta}-phase SiC peak features at reflection angles of 35.68\textdegree, 60.04\textdegree, and 71.84\textdegree for the 1\%, 5\%, and 10\% conditions. Scherrer's first peak approximation (corrected for instrumental broadening by a LaB6 standard) on the SiC (111) peak yields a consistent crystallite value of 8 nm for all three samples. This agrees with particle size statistics from TEM and suggests that the SiC inclusions do not grow or sinter during densification. Scherrer's first peak approximation on the Si (111) peak revealed crystallite sizes of 147 nm, 115 nm, 88 nm, and 70 nm for 0\%, 1\%, 5\%, and 10\% respectively. The decrease in crystallite size as a function of SiC volume fraction is consistent with the reduced densification kinetics discussed earlier, implying that the addition of SiC inclusions slows down grain growth and densification.

Figures~\ref{fig:exp}(a) and~\ref{fig:exp}(b) show a cross-section of the sintered sample with 5\% volume fraction of SiC inclusions. The sample was prepared by focused ion beam milling (FIB) and characterized by scanning tunneling electron microscopy (STEM) and energy-dispersive X-ray spectroscopy (EDS) on a Titan Themis 300. Domains with a size around 100 nm are clearly distinguishable, consistent with the crystal size for Si obtained by the analysis of the XRD diffraction patterns. Spherical inclusions around 8 nm in size are also clearly present. The corresponding elemental map suggests that these small inclusions are carbon-rich. The elemental analysis and the fact that the size is consistent with the XRD results and with the TEM in Fig.~\ref{fig:tem}(a) strongly suggests that the small inclusions are the SiC nanocrystals. Some voids are also present, again the likely result of reduced densification rate for the samples with inclusions.

%% file: Sections/thermoelectric_transport_properties.tex
\section{THERMOELECTRIC TRANSPORT PROPERTIES}

\begin{figure*}[t]
\begin{center}
\includegraphics[width=1\textwidth]{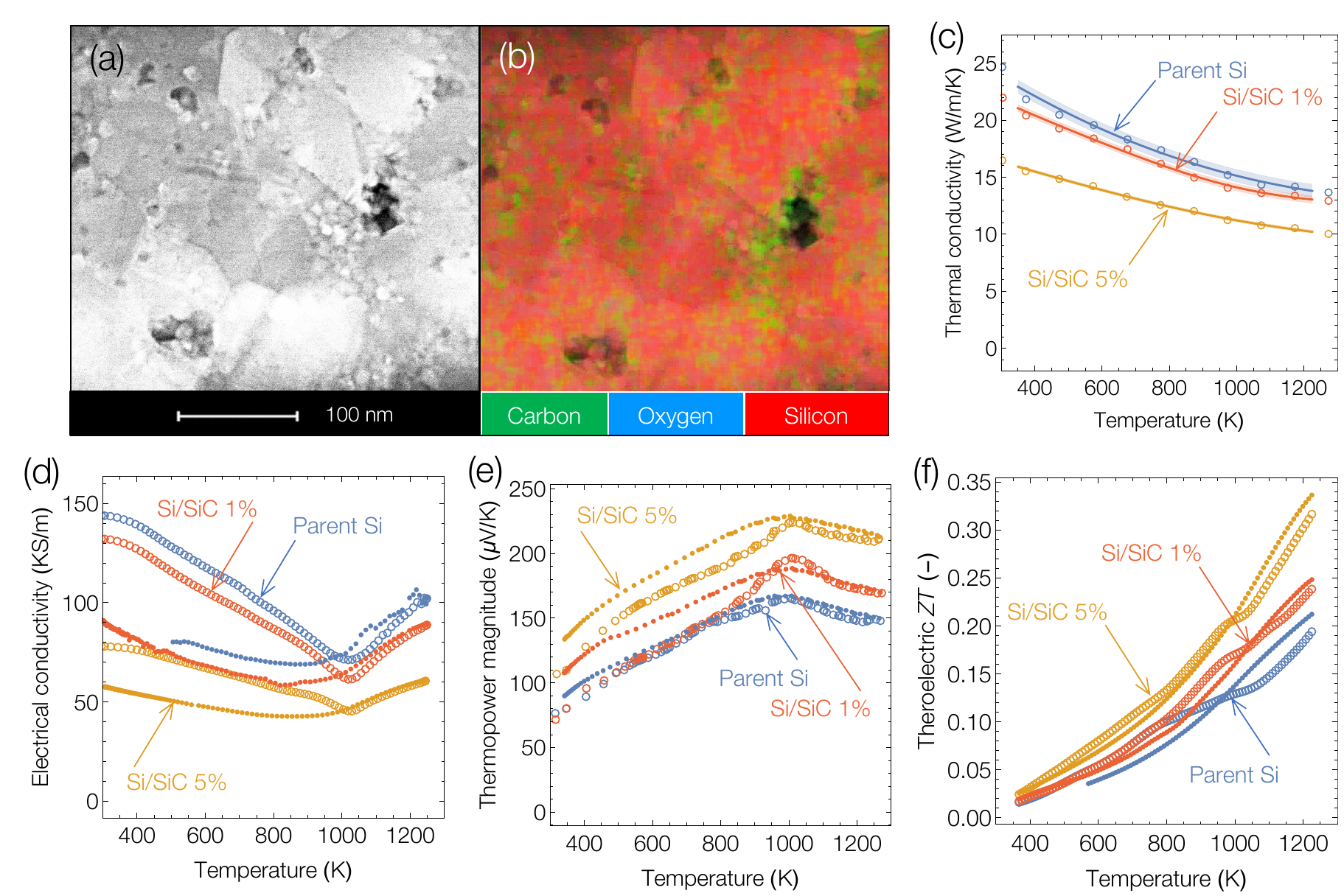}
\end{center}
\caption{Experimental characterization of the sintered SiC/Si nanocomposites. Panel (a) shows a TEM micrograph of a fibbed cross-section of the SiC/Si composite with a 5\% volume fraction of SiC particles. Image (b) shows the corresponding composition map obtained from energy-dispersive X-ray spectroscopy (EDS). Plots (c-f) show the measured transport properties as a function of the temperature of the parent silicon with no inclusions (blue) and SiC/Si nanocomposites with 1\% (red) and 5\% (gold) of SiC nanoparticles by volume. Plot (c) shows the total thermal conductivity. Plot (d\&e) show the materials’ experimentally measured electrical conductivity $\sigma$, and Seebeck coefficient $S$, respectively. In these plots, the open circles show the properties during initial heating from room temperature, and the closed dots show the properties measured as the system is cooled back down to room temperature. The non-monotonic trends, and the divergence of the heating and cooling curves in $\sigma$ and $S$ are due to changes in dissolved electrically active P as described in the text. Overall, it can be seen that the addition of inclusions degrades the electrical conductivity but increases the magnitude of the Seebeck coefficient sufficiently that the overall power factor is increased. Panel (f) shows the thermoelectric figure of merit $ZT$ computed using Gaussian process (GP) regression fit to the experimental data in (c--d). Again open circles are for heating, and dots for cooling. The uncertainty band from the GP model is a similar width to the plotting symbols and so if not drawn for clarity.}
\label{fig:exp}
\end{figure*}

The electrical and thermal transport properties of the samples described above were characterized at the Jet Propulsion Laboratory, using commercially available equipment for thermal diffusivity and dedicated custom equipment for carrier concentration, electrical conductivity, and Seebeck coefficient. Thermal conductivity was calculated from thermal diffusivity measured using a commercial Netzch LFA 404 Laser Flash Analysis (LFA), and density measured by Archimedes method and literature values for heat capacity and thermal expansion~\cite{swenson1983recommended, shanks1963thermal}. The measured thermal conductivity is shown in Fig.~\ref{fig:exp}(c). Even without SiC inclusion,  at room temperature, the thermal conductivity of the heavily doped polycrystalline Si is less than one-fifth of single-crystal Si. The thermal conductivity is dominated by phonon transport, and the addition of nanoparticles further reduces the thermal conductivity.

All electrical property measurements were made during a temperature sweep where the sample was heated from room temperature to 1300 K at a rate of 180 K/hour, and then cooled back down to room temperature. Electrical conductivity and carrier concentration were measured from a custom high-temperature Hall Effect using a four-point probe Van der Pauw method described in~\cite{borup2012measurement}. The Seebeck coefficient was measured using a small {\textDelta}T Seebeck coefficient measurement system with the procedure described in~\cite{wood1985measurement}. 

The measured electrical conductivity is plotted in Fig.~\ref{fig:exp}(d). These show that on initial heating, conductivity falls linearly with temperature up to around 900 K, after which it rises abruptly. On cooling, the conductivity does not follow the same return path. Corresponding behavior is seen in the measured carrier concentrations (described in detail later) indicating that this irreversible change arises from an initial supersaturation of electrically active P dopant in the as-sintered samples. Above around 900 K the dissolved P becomes mobile and drops out of solution (or forms defect clusters that are not electron-donating), lowering the carrier concentration. However, the solubility of P rises with temperature, and above around 1000 K dopant is re-dissolved into solution, and the carrier concentration and conductivity rise. During the cooling process, the P dopant drops out of the solution until around 800 K, below which the P is once more immobile. In addition to the change in conductivity with temperature, there is a clear trend for the reduction in conductivity with increasing volume fraction of SiC inclusions due to the extra electron scattering that they cause. 

The irreversible change in carrier concentration is also seen in the thermopower, the \emph{magnitude} of which is plotted in Fig.~\ref{fig:exp}(e).  In parabolic band materials with energy independent electron scattering the Seebeck coefficient has a $N_i^{-2/3}$ dependence on the carrier density $N_i$~\cite{snyder2011complex}, so in this case, for each sample, the thermopower increases moderately after the heating-cooling cycle due to the removal of the initial P supersaturation. However, a more significant increase in thermopower is seen to come from the addition of SiC nanoinclusions. We note that the addition of undoped SiC inclusions has the effect of reducing the overall doping concentration in the material -- an effect that would cause the magnitude of the Seebeck coefficient to increase and the electrical conductivity to decrease, consistent with the trends seen in Fig.~\ref{fig:exp}. With 5\% of the P-doped Si replaced with SiC one would expect a 5\% reduction in the net carrier concentration assuming that the Si matrix maintained the same P concentration. However, the room temperature carrier concentrations measured in the as-sintered 0\% and 5\% SiC samples are 3.3$\times$10\textsuperscript{20} 1/cm\textsuperscript{3} and 2.9$\times$10\textsuperscript{20}, respectively. This 12\% decrease in the carrier concentration perhaps indicates that P is segregating to grain boundaries or the Si/SiC interface. More importantly, although larger than expected, a 12\% decrease in carrier concentration is expected to yield only a $\sim$9\% increase in Seebeck coefficient, not the $\sim$50\% increase that was measured. This strongly suggests that the reduction in carrier concentration alone is not sufficient to explain the magnitude of the change in electrical transport properties. We hypothesize that the additional enhancement in thermopower comes from electron energy filtering due to scattering from inclusions. 

To examine this hypothesis, we have developed a semiclassical model of electron transport that takes as its only input the measured carrier concentration. This model is able to reproduce the measured variation in electrical transport properties with temperature, provided that we account for the scattering of electrons by  SiC nanoinclusions. The model shows that this scattering imparts an electron energy filtering effect that can \emph{increase} thermopower --- this enhancement is significant, almost a 100\% increase in the room temperature Seebeck coefficient from the addition of 5\% volume fraction of SiC.

The measurements of $\sigma$ and $S$ in Fig.~\ref{fig:exp}(c)--(d) were made on the same sample during the same temperature sweep, but the measurements were not synchronous, so to compute the thermoelectric figure of merit $ZT$ (plotted in Fig.~\ref{fig:exp}(f)) the experimentally measured $\sigma$, $S$, and $\kappa$ were interpolated using Gaussian process regression. Overall, the parent sample (0\% SiC) has somewhat poor thermoelectric performance (ZT = 0.19 at 1,300 K). However, the addition of the SiC inclusions has a significant effect on the transport properties, with 5\% of SiC inclusions producing roughly a 50\% increase in $ZT$. Most of this increase stems from the reduction in thermal conductivity, with the reduction of the electrical conductivity due to inclusions compensated by an increase in the Seebeck coefficient that produces a small overall increase in the power factor. It can also be seen that the irreversible carrier concentration change from heating to cooling creates a kink in the $ZT$ profiles during heating that occurs at the same temperature for all three samples. Although there is some loss in $ZT$ below 800 K after the first heating cycle, the additions of the nanoparticles provide a large enhancement in $ZT$ that remains after heating and cooling. The irreversible change in carrier concentration does not affect the thermoelectric performance above 1000 K, which is the targeted operating regime for these materials, but it is opportune for validation of our transport model and it means that we have six separate data sets in which the carrier concentration is varied separately from the SiC fraction.

The concept of electron energy filtering for the enhancement of power factor relies on the use of energy-selective electron scattering to impede the transport of electrons with low energy while leaving unimpeded electrons in energy states with occupancy most sensitive to changes in temperature. In the \textit{n}-type Si studied in this work the filtering is provided by the band offset of the SiC nanoparticles which produces a local step in the conduction band edge, as is shown schematically in Fig.~\ref{fig:cc}(a). The inclusions present a barrier to the propagation of low-energy electrons --- and in principle, one can engineer the height of this barrier by doping the nanoinclusions, or by substituting the SiC for a material with a different bandgap. The concept of electron filtering is described well in other works~\cite{gayner2020energy, guan2021enhancement, pham2021improved, PhysRevB.89.075204, PhysRevB.87.075204, schmidt2010thermal, chen2005nanoscale, PhysRevB.68.125210}, but we reiterate it here mathematically to set the scene for the model developed in the next section, and to provide the necessary background to our exploration of the theoretical limits energy filtering at the end of this article.

%% file: Sections/model_transport_coefficients_in_bulk_thermoelectrics.tex
\subsection{Model of Transport Coefficients in Bulk Thermoelectrics}

The electrical conductivity and thermopower of a population of independent charge carriers can be derived from the Boltzmann transport equation by integrating the contribution from all carriers states. In an isotropic system where the states can be enumerated by their energy, and using the single relaxation time approximation for the collision operator, these can be written as integrals over the carrier energy $E$, so that $\sigma$, $S$, and $\kappa_e$ are given by~\cite{chen2005nanoscale}
\begin{equation}\label{eq:sigma}
    \sigma = \dfrac{-1}{3}e^2\int\chi(E,T)\tau(E,T)dE=\dfrac{-1}{3}e^2\Delta_0,
\end{equation}
\begin{equation}\label{eq:s}
    S = \dfrac{-1}{eT}\dfrac{\int\gamma(E,T)\tau(E,T)dE}{\int\chi(E,T)\tau(E,T)dE}=\dfrac{-1}{eT}(\Delta_1-E_f),
\end{equation}
\begin{equation}\label{eq:ke}
    \kappa_e = \dfrac{-1}{3T}e^2 \left(\int\zeta(E,T) \tau(E,T) dE -\dfrac{(\int\gamma(E,T) \tau(E,T) dE)^2}{\int\chi(E,T)\tau(E,T) dE}\right) \\ =\dfrac{-1}{3T}\Delta_0(\Delta_2-\Delta_1^2).
\end{equation}
Here the function $\chi(E,T)=\nu^2(E)D(E)\dfrac{df(E,E_f,T)}{dE}$, lumps together the material's density of carrier states, $D(E)$, and group velocity, $\nu(E)$, with the energy derivative of the Fermi-Dirac occupancy, $f(E,E_f,T)$, where $E_f$ is the Fermi level. The functions $\gamma(E,T)=(E-E_f)\chi(E,T)$ and $\zeta(E,T)=(E-E_f)^2 \chi(E,T)$. Eqs.~\ref{eq:sigma},~\ref{eq:s} and~\ref{eq:ke} also express the relationship between the transport properties and $\Delta_n$, the moments of the distribution of conductivity over carriers with different energy, defined as
\begin{equation}\label{eq:delta_n}
\Delta_n =\left\{\begin{matrix}
\int \chi\tau dE & n=0\\ 
\frac{1}{\Delta_0}\int E^n \chi \tau dE & n\neq 0
\end{matrix}\right.
\end{equation}
The Seebeck coefficient from Eq.~\ref{eq:s} obtains its largest magnitude by maximizing the asymmetry of product $D \tau \nu^2$ about the Fermi level to move its center of current, $\Delta_1$, away from the Fermi level. In bulk semiconductors, the relaxation time, $\tau$ -- from ionic impurities and phonons -- is inversely proportional to the electronic density of states, $D(E)$~\cite{lundstrom2012near}, with a weak dependence on carrier energy and a prefactor that appears in the numerator and denominator of Eq.~\ref{eq:s}. This makes the Seebeck coefficient quite insensitive to the overall magnitude of electron scattering rate and leaves $\nu(E)$ as the only property that impacts the Seebeck coefficient. As a result, most approaches to optimize power factor, $\sigma S^2$, focus on engineering the band structure and Fermi level to tune the charge carriers concentration and effective mass to align the Fermi energy to where the density of states is changing most rapidly~\cite{peienergy}. Unfortunately, these parameters produce countervailing responses in $S$ and $\sigma$, so the overall scope for enhancing the power factor is limited.

\begin{figure*}[t]
\begin{center}
\includegraphics[width=0.90\textwidth]{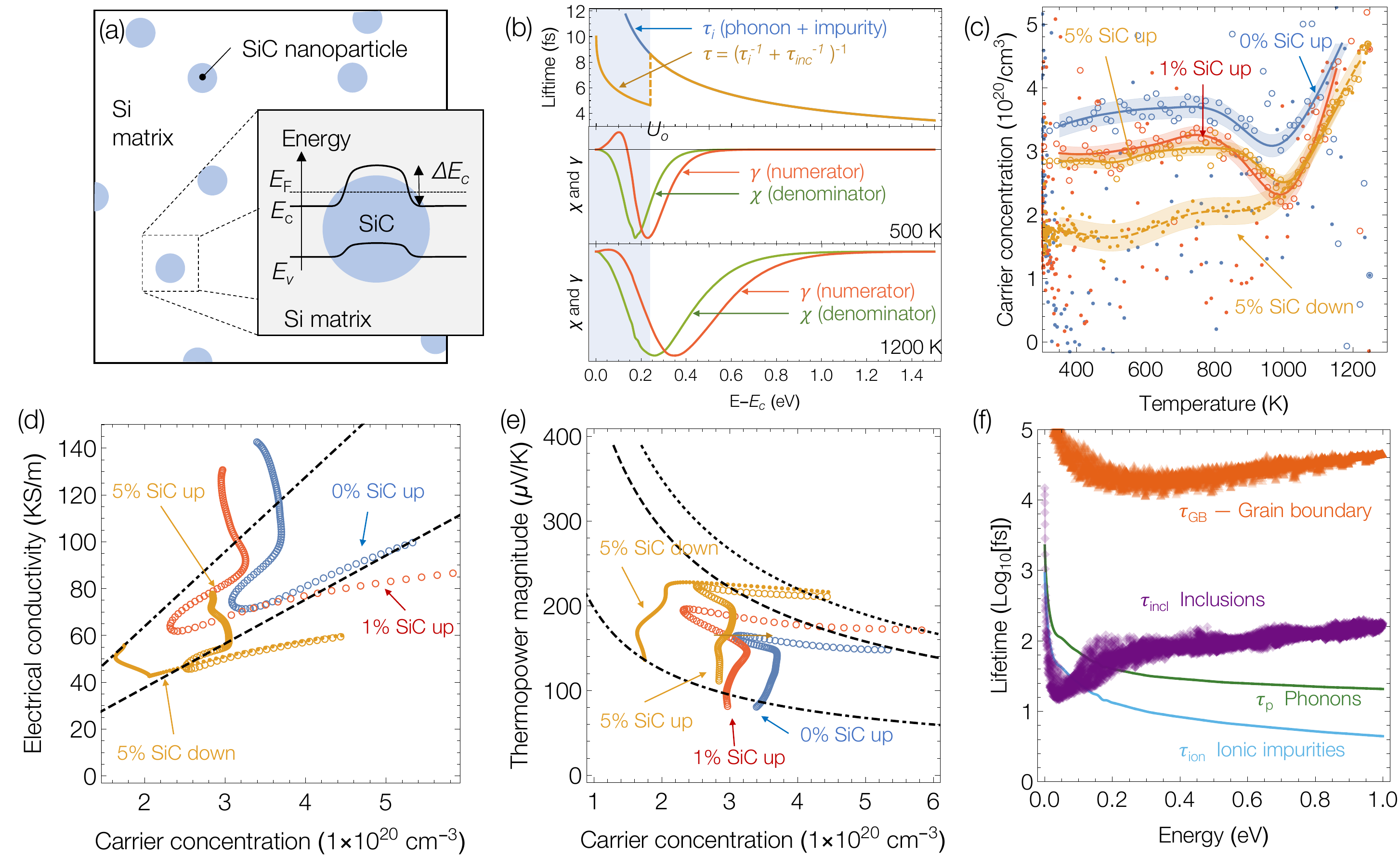}
\end{center}
\caption{(a) Schematic picture of the energy barrier $\Delta E_c$ for filtering conduction band electrons in Si due to the conduction band offset of embedded SiC nanoparticles. Panel (b) illustrates the electron energy filtering concept. The upper plot shows the energy dependence of the electron scattering time, including an additional filtering scattering process that is felt by all electrons with energy less than $U_o$. The lower plots show the kernels $\chi$ and $\gamma$, normalized, and plotted at 500 K (middle) and 1200 K (bottom). The additional electron filtering scattering in the shaded region causes a larger reduction of the $\tau$ weighted integral of $\chi$ than $\gamma$. Panel (c) shows the temperature dependence of the experimentally measured carrier concentration, $N_i$, for P-doped silicon with 0\% (blue), 1\% (red) and 5\% (gold) volume fraction of SiC. You open circles and dots show the measurements made during heating and cooling respectively.  The solid and dashed lines show Gaussian process regression fit (and corresponding uncertainty) to the four least noisy data sets. Plots (d\&e) show the variation in electrical conductivity and thermopower with carrier concentration using the same color coding and symbols as in (c). In (d) the black lines provide a guide to the eye to illustrate the isothermal change in conductivity due to differences in carrier concentration. The dashed line compares how the 0\% SiC conductivity at 1285 K would change with $N_i$ and it can be seen that the materials with SiC inclusions are more restive than a material with the same $N_i$ but without inclusions. The dot-dashed line shows a similar guide fit to the 300 K conductivity of the material with 5\% SiC after heating and cooling. In plot (e) the black lines show isothermal $N_i^{-2/3}$ dependence expected of parabolic band material. As in (d), the dashed and dot-dashed lines are fit to the high-temperature measurement of the material with 0\% SiC, and the low-temperature measurement in the cooled material with 5\% SiC. The dotted line is fit to the high-temperature measurement of the material with 1\% SiC. At both high and low temperatures the materials with a higher fraction of inclusions have thermopower larger than that predicted by the variation in $N_i$. Panel (f) plots the electron lifetime for the different scattering mechanisms described in the text for Si at 500 K with a carrier population of $\mathrm{2.8 \times 10^{20}\ 1/cm^3}$.  It can be seen that the electron-inclusion scattering (purple) is dominant for electrons with energies less than $\sim$0.12 eV. The scattering of electrons with higher energies is dominated by electron-impurity scattering (blue). The electron-grain boundary ($l_g= 50$ nm) and electron-inclusion (inclusion radius of $\mathrm{4\ nm}$) for 5\% SiC inclusion are two additional scattering terms Si nanocomposite and are shown in orange and green, respectively.}
\label{fig:cc}
\end{figure*}

An alternative strategy for generating asymmetry in $D\tau \nu^2$ is to add extrinsic scattering processes (a task that is easier than engineering intrinsic properties) to break the reciprocity of $D$ and $\tau$. Introducing any new scattering mechanism shortens the electron relaxation time and hence reduces $\sigma$. For the Seebeck coefficient, however, $\tau$ appears in both numerator and denominator of Eq.~\ref{eq:s}, consequently both numerator and denominator are decreased by the additional scattering mechanism. The central strategy of electron energy filtering is to introduce a scattering mechanism that can reduce the denominator of $S$ faster than the numerator so that the overall magnitude of the Seebeck coefficient is increased. 

The electron energy filtering concept is illustrated graphically in Fig.~\ref{fig:cc}(b), which plots normalized $\chi$ and $\gamma$ functions for the conduction band of silicon on the same axes as $\tau$, the electron lifetime. For \textit{n}-doped semiconductors, $\chi$ is negative for any $E$ in the conduction band, and thus $\gamma$ is positive for electrons with energy lower than $E_f$. The high energy tails in $\chi$ and $\gamma$ look functionally similar, and imposing electron scattering in the high energy region has a similar scaling effect on the magnitude of the integrals in the numerator and denominator of Eq.~\ref{eq:s}. At low energies around and below the Fermi energy, $\chi$ and $\gamma$ are functionally very different. In this region, $\gamma$ contributes little or negatively to its integrated magnitude, whereas $\chi$ contributes strongly to its integrated magnitude. Hence introducing processes for selectively scattering electrons with energies in the blue shaded region of Fig.~\ref{fig:cc}(b) will decrease the denominator of Eq.~\ref{eq:s} faster than the numerator, yielding an increase in $S$ -- and can even increase the power factor.

The addition of SiC dispersoids to Si provides additional electron and phonon scattering centers that could enable electron filtering; however, since phosphorous is not expected to dissolve in SiC, the presence of the nanoinclusions effectively reduces the overall carrier concentration. To discriminate the effects of changes in carrier concentration from electron filtering we require a self-consistent and quantitatively accurate model of the electrical transport properties. To this end, we have developed a semiclassical transport model that computed Eqs.~\ref{eq:sigma} and~\ref{eq:s} using the ab initio computed band structure of pure Si in combination with the experimentally measured carrier concentration. The calculations were performed using a python package,~\emph{thermoelectric.py}, that we have made available for download through GitHub~\cite{hosseini2019}.

\subsection{Calculation of Intrinsic Properties}

The terms $D(E)$, and $\nu(E)$, in functions $\chi$ and $\gamma$ in Eqs.~\ref{eq:sigma}--\ref{eq:delta_n} were derived from the conduction band of Si computed with density functional theory (DFT) using the Vienna Ab initio Simulation Package (VASP) \cite{PhysRevB.54.11169, kresse1996phys, PhysRevB.49.14251, kresse1993ab}. The calculations were performed using the generalized gradient approximation (GGA) with the Perdew-Burke-Ernzerhof exchange correlation functional (PBE)~\cite{perdew1996generalized}. Ion cores were represented with projector augmented wave (PAW) pseudopotentials~\cite{kresse1999ultrasoft, blochl1994projector}, the Kohn-Sham wave functions constructed using a planewave basis set with a 700 eV energy cutoff, and a $\mathrm{12\times12\times12}$ Monkhorst-Pack k-point grid was used to sample the Brillouin zone~\cite{monkhorst1976special}. The Si primitive cell and atomic basis were relaxed to minimize forces on the atoms to better than 10\textsuperscript{-6} eV/\AA. The electronic band structure used to compute $D(E)$ and $\nu(E)$ were generated from a $\mathrm{45\times45\times45}$ k-point grid. These intrinsic materials' properties were treated as temperature independent. The group velocity was obtained from the conduction band curvature, $\nu = 1/\hbar |\nabla_k E|$ along the $\mathrm{\left\langle 100\right\rangle}$ direction on the $\Gamma$ to $X$ Brillouin zone path.

\subsection{Evaluation of Carrier Concentration and Fermi Level}

The final term that appears in the distributions $\chi$ and $\gamma$ is the Fermi level $E_f$. This is not an intrinsic property, $E_f$ is strongly dependent on the carrier concentration, $N_i$, and experimentally this is found to vary non-monotonically with temperature as the solubility of the phosphorus dopant changes. Rather than model the physics of the temperature dependence of carrier concentration (which is unrelated to electron transport), we use the empirically measured carrier concentration as an input and then compute the Fermi level that gives the same carrier population in the DFT computed conduction band. This circumvents the problem that DFT underestimates the bandgap as the Fermi level is computed self-consistently from the conduction band using the conduction band edge to set the reference frame. The carrier concentration was measured concurrently with $\sigma$ and $S$ during the temperature, but the stability of the measurements was inconsistent as can be seen in Fig.~\ref{fig:cc}(c). In some portions of the experiment the measurements had little noise and showed a smooth trend, while in others the measurements became wildly noisy producing nonsensical values. Smooth interpolation of these data was obtained by first eliminating the extreme (nonphysical) outliers and performing Gaussian process regression using a kernel constructed from the sum of white noise and Mat\'erns 5/2 covariance function. With this approach, reasonable fits for $N_i$ with moderate uncertainty were obtained for the heating sweep of all three materials, but only for the cooling sweep of the material with 5\% SiC, (which are plotted with solid and dashed lines in Fig.~\ref{fig:cc}(c)). The noise in the experimental values for $N_i$ was too great to provide a meaningful fit to the cooling sweep for the materials with 0 \% and 1\% SiC, and so only the four reliable $N_i$ data sets were used in the modeling that follows.

The Fermi levels, $E_f$, at a given $N_i$ was determined by inverting the relationship
\begin{equation}
    N_i = \int_0^\infty dE D(E) f\left(E,E_f,T\right).
\end{equation}
This inversion was achieved numerically by generating a table of the carrier concentration $N_i$ computed over a range of $T$ and $E_f$ and then using this table to interpolate $E_f$ at a given value carrier concentration. The Fermi levels computed for the carrier concentrations in Fig.~\ref{fig:cc}(c) were in the conduction band at all temperatures. This implies that there is a range of electron energies for which $\gamma$ is positive. The addition of a scattering mechanism for low-energy electrons is thus expected to lead to an increase in the magnitude of $S$.

\subsection{Models of Electron Scattering Processes}

Semiconductor TEs are generally doped to beyond their saturation level (supersaturate solutions). In these materials, strongly screened Columbic force induced by ionized impurities provide the main source of scattering. The transition rate between initial ($E_i$) and final ($E_f$) energy states has $SR(E_i,E_f )=\frac{2\pi N_i e^4 L_D^4}{(4\pi \epsilon \epsilon_o)^2 \hbar \Omega} \delta(E_f-E_i)$, where $L_c$, $\Omega$, $\epsilon$, and $\epsilon_o$ are the Debye length, volume, relative and vacuum permittivity, respectively~\cite{lundstrom2009fundamentals}. In this case, the electron lifetime is defined as~\cite{lundstrom2009fundamentals}
\begin{equation}\label{eq:tauim}
 \tau_{ion}(E)=\frac{\hbar}{\pi N_i \left(\frac{e^2 L_D^2}{4\pi \epsilon \epsilon_o}\right)^2 D(E)}.
\end{equation}
We use $\epsilon = 11.7$ to model permittivity in Si~\cite{levinshtein2001properties}. The Debye length has generalized form of~\cite{mondal1986effect}
\begin{equation}\label{eq:ld}
L_D=\frac{e^2 N_c}{4 \pi \epsilon \epsilon_o k_B T} \left [F_{-\frac{1}{2}}(\eta)+\frac{15 \alpha k_B T}{4} F_{\frac{1}{2}}(\eta)\right],
\end{equation}
where $N_c=2\left(\frac{m_c k_B T}{2\pi \hbar^2} \right)^{\frac{3}{2}}$ is the effective density of states in the conduction band, $m_c$ is the conduction band effective mass and $\eta = E_f/k_BT$. While the electron lifetime in Eq.~\ref{eq:tauim} serves reasonably well for many semiconductors it has two shortcomings. The Born approximation fails for slow moving electrons in a Coulombic potential, and modeling the scattering as the colligative effect of multiple  single impurity centers fails to capture any interference effects that arise as the electron wave function propagate through a random distribution of impurities. We modeled the variation in the conduction band effective mass with temperature using $m_c (T)=m_c^* (1+5 \alpha k_B T)$~\cite{riffe2002temperature}, with $m^*_c = 0.23m_e$, where $m_e$ is free electron rest mass ($9.11\times10^{-31}$ kg). The reciprocal energy, $\alpha = 0.5$ eV\textsuperscript{-1} describes the deviation of the conduction band from parabolic. This model assumes linear dependency on temperature and does not count for degeneracy in high carrier population. A better model that captures the effects from dopant concentration needs further study. At moderate and low carrier populations, the electron-ion charged scattering rate is modeled using Brooks and Herring expression~\cite{lundstrom2009fundamentals}
\begin{equation}\label{eq:tauim2}
\tau_{im}= \frac{16\pi\sqrt{2m_c}(4\pi \epsilon \epsilon_o)^2} {e^4 N_i\left(\ln \left( 1+\beta \right) - \frac{\beta}{1+\beta} \right)} E^{\frac{3}{2}},
\end{equation}
where $\beta = \frac{8m_c E L_D^2}{\hbar^2}$. For more details on the derivation of Eqs.~\ref{eq:tauim} and ~\ref{eq:tauim2}, see Ref.~\cite{lundstrom2009fundamentals}.

The second important scattering mechanism in nonpolar semiconductors like Si is due to  scattering from acoustic phonons --- it is particularly important at high temperature. Ravich has derived an expression for the electron scattering lifetime due to this inelastic process~\cite{ravich1971scattering}
\begin{equation}\label{eq:taup}
\tau_p(E)=\frac{\rho \nu_s^2 \hbar}{\pi \Phi_A^2 k_B T D(E)} \left ( \left[1-\frac{\alpha E}{1+2\alpha E} \left(1-\frac{\Phi_v}{\Phi_A} \right)\right]^2-\frac{8}{3} \frac{\alpha E(1+ \alpha E)}{(1+2 \alpha E)^2}\frac{D_v}{D_A} \right)^{-1}.
\end{equation}
Here $\rho$ and $\nu_s$ are the crystal’s mass density and speed of sound, respectively. The terms $\Phi_v$ and $\Phi_A$ are the electron and hole deformation potentials, which equal 2.94 eV and 9.5 eV, respectively~\cite{75176}. We use $\rho = 2329$ kg/m\textsuperscript{3} and $\nu_s=\sqrt(B/\rho)$, where the bulk modulus $B=98$ GPA~\cite{levinshtein2001properties}. This equation accounts for both absorption and emission of phonons. 

The variation of $\tau_\mathrm{ion}$ and $\tau_p$ with electron energy are plotted in Fig.~\ref{fig:cc}(f). Over the range of temperatures and carrier concentrations studied here $\tau_\mathrm{ion}$ dominates except for very high temperatures where $\tau_p$ takes precedence over $\tau_{ion}$. Other intrinsic scattering processes such as electron-electron and electron intervalley scattering are negligible compared to these and so are ignored.

%% file: Sections/model_transport_in_nanocomposite_thermoelectrics.tex
\section{MODEL OF TRANSPORT IN NANOCOMPOSITE THERMOELECTRICS}

There are three significant differences between the material with and without silicon carbide inclusions that the electrical transport properties: (1) There are differences between the doping concentration and resulting charge carrier concentration and Fermi energy. (2) The additional inclusions provide an extra source of electron scattering (with lifetime $\tau_{np}$) that is not present in the parent Si. (3) The grain size differences between the three materials and thus the rate of electron scattering from grain boundaries changes (which has lifetime $\tau_{gb}$). To compute the electron lifetime from the two extra scattering processes we used Fermi’s golden rule to relate the transmission probability from the initial energy state to the distribution of final energy states for a given time-invariant potential. In the case where energy conservation is imposed (elastic scattering) the scattering rate in Born approximation can be written as~\cite{hosseini2021enhanced, lee2010effects}
\begin{equation}\label{eq:fg}
    \tau^{-1}(s) = \frac{N}{(2\pi)^2\hbar}\int_{E(k')=0}\frac{M_{kk'}\overline{M}_{kk'}}{\nabla E(k')}(1-\cos\theta)dS(k').
\end{equation}
Here, $M_{kk'}$ is the matrix element operator shows the coupling strength between initial and final wavefunctions and the number of ways the transmission may happen, $N$ is the number density of scattering source and $\theta$ is the angle through which the electron's momentum is turned between the initial and scattered states. For Bloch waves, $M_{kk'}$ is defined as $M_{kk'}= \int e^{i(k'-k).r} U(r)dr$~\cite{nag2012electron}. In Eq.~\ref{eq:fg}, $S(k')$ represents isoenergic surface of electronic states in $k$-space. For semiconductors with Si-like  band structures with indirect degenerate band gap, the contour surfaces of k-states around the conduction band valley with energy $E(k)$ above the conduction band edge is approximated well by an ellipsoid $E(k)=\hbar^2 \left[\frac{(k_l-k_{ol} )^2}{2m_l^*} +\frac{k_t^2}{m_t^*}\right]$, where $k_l$ and $k_t $ are the components of the wavevector that are parallel ant transverse to the long axis of the conduction band valley. The term $k_{ol}$ describes the location of the conduction band minimum, while $m_l^*$ and $m_t^*$ are the effective masses of electrons traveling along and transverse to the conduction band valley, respectively. For silicon, $m_l^*=0.98m_o$ and $m_t^*=0.19m_o$ where $m_o$ is free electron rest mass, and $k_{ol}=0.85 2\pi/a$ where $a$ is silicon's lattice parameter of 0.543 nm~\cite{levinshtein2001properties}.

\subsection{Model of Electron Lifetime for Scattering by Nanoparticles}

The band alignment at the interface of nanoparticles presents a barrier to electron transport equal to the conduction band offset, $\Delta\!E_c$ between bulk silicon and the inclusions, as is shown in Fig.~\ref{fig:cc}(a). 
For spherical nanoparticles, the scattering potential term, given as, $U(r)=\Delta\!E_c \Pi(r_o-r)$, where $r_o$ is the nanoparticle’s radius and $\Pi(r)$ is a dimensionless boxcar function equal to unit inside and zero outside of the particles. For the spherical symmetric potential, $M_{kk'}$ only depends on $\bm{q}=\bm{k}'-\bm{k}$ and is defined as~\cite{hosseini2021mitigating, hosseini2021prediction}
\begin{equation}\label{eq:m}
    M_{kk'}=\frac{4\pi \Delta\!E_c}{|\bm{q}|^2}\left( \frac{1}{|\bm{q}|}\sin\left(r_o|\bm{q}|\right)-r_o\cos\left(r_o|\bm{q}|\right)\right).
\end{equation}

At equilibrium, the Fermi energy level of nanoparticles and parent material aligned leaving the band offset between SiC nanoparticles and silicon, $\Delta\!E_c$, equal to the difference between Fermi energy level and conduction band edge of the SiC. For intrinsic semiconductors Fermi energy level is located at the middle of band gap so that $\Delta\!E_c=\frac{1}{2}E_g$. The SiC band gap varies from 2.36 eV at 300 K down to 2.036 eV at 1200 K following ($E_g  = 2.39-6.0\times10^{-4}\times \frac{T^2}{T+1200}$)~\cite{levinshtein2001properties}. Such a variation has negligible effect on scattering rate so that we used temperature independent value of $E_g$ =2.19 eV (and therefore $\Delta\!E_c = 1.095$ eV) to model electron-nanoparticle scattering rate. Note that $N$ in Eq.~\ref{eq:fg}, is the number density of nanoparticles and is equal to $N=\frac{3\phi}{4\pi r_o^3}$, with $\phi$ the volume fraction of nanoparticle. We have computed the rates of electron scattering from SiC nanoparticles by using Eq.~\ref{eq:m} in~\ref{eq:fg} and integrating over the ellipsoidal approximation for the isoenergetic surfaces. The resulting distribution of scattering times $\tau_{inc}$ is shown in purple in Fig.~\ref{fig:cc}(f).

\subsection{Model of Electron Lifetime for Scattering by Grain Boundaries}

Along with the change in dopant concentration, the addition of 1\% and 5\% of SiC nanoparticles results in a 22\% and 40\% reduction in the grain size, respectively. 
It is known that grain boundaries can cause an electron filtering effect, particularly if the boundaries include segregated species such as oxygen that provide centers for trapping charge carriers~\cite{martin2009enhanced}. However, this effect only becomes significant in much smaller grain sizes. For our Si/SiC nanocomposites, even with a 40\% size reduction, the grains are still an order of magnitude larger than the average electron mean free path in P-doped Si (which is only a few nanometers only at room temperature for carrier concentrations in excess of 10\textsuperscript{20} 1/cm\textsuperscript{3})~\cite{bux2009nanostructured}. Furthermore, we have computed the rate of electron scattering from grains (this is of special importance in the next section where we evaluate the scope of enhancement in power factor in Si nanocomposites) using the approach by Minnich et al. in Ref.~\cite{PhysRevB.80.155327} in which they have modeled grain boundaries as decomposition of many local regions, each interacts independently with charge carriers and coherently scatters electron waves. The model potential for grain boundaries in their work described as
\begin{equation}\label{eq:ugb}
    U_{GB} =\left\{\begin{matrix}
 U_g e^{\frac{-|z|}{z_o}}& r<r_{GB} \\ 
 0& r>r_{GB}
\end{matrix}\right.
\end{equation}
In this equation, $z$ is the direction normal to the grain with $z=0$ at the center of the grain boundary, $r_{GB}$ is a constant on the order of the screening length, and $z_o$ is a constant related to the size of the depletion region. $U_g$ in this model is proposed as, $U_g=\frac{e^2 N_t^2}{8 \epsilon \epsilon_o N_c}$, where $\epsilon$ and $\epsilon_o$ are relative and vacuum permittivity, respectively, $N_c$ is the doping concentration, and $N_t$ is the area density of traps. The matrix element of this potential is
\begin{equation}\label{eq:mg}
    M_{kk'}=4\pi U_g \left[ \frac{z_o}{1+(q_zz_o)^2} \right]r_o^2\left[ \frac{J_1(q_rr_o)}{q_rr_o} \right]
\end{equation}
where $J_1 (q_r r_o )$ is the first-order Bessel function of the first kind, $q=k-k'$, $q_r$ and $q_z$ are the $r$ and $z$ component of $q$, respectively. Equations~\ref{eq:fg} and~\ref{eq:mg} are used to compute $\tau_{gb}$. Unfortunately, there is a limit information about the trap area density ($N_t$) and the exact value of $z_o$ and $r_o$. Nevertheless, we know that depletion regime and the screening length are on the order of few nm. We used $N_t  = 10^{13}  \mathrm{\frac{1}{cm^2}}$ for trap density of doped silicon, $z_o=1$ nm and $r_o=1$ nm~\cite{minnich2009modeling}.
 
\subsection{Correction to the Electron Density of State in Nanocomposites}

The Fermi level is set by the local doping and resulting carrier concentration in the silicon matrix. However, the carrier concentration that is measured experimentally is the average carrier concentration, which for the nanocomposites will be the carrier concentration in the Si matrix diluted by the volume of inclusions embedded in the matrix. This has the effect of reducing the density of electronic states which impacts the conductivity. Thus to predict $\sigma$ and $S$ for a nanocomposite from an experimentally measured carrier density $N_i$, we first adjust the carrier density up to obtain the concentration in the Si matrix $N_{i~ \mathrm{matrix}}=N_{i~\mathrm{measured}}/\left(1-\phi\right)$ and use this to find the Fermi level. Then, after computing $\sigma$ and $S$, the conductivity is adjusted down to give the effective conductivity $\sigma_{\mathrm{eff}}= (1 - \phi)\sigma$ which accounts for the reduced density of states. This correction is not required for the Seebeck coefficient since the changes in the density of state cancel out for the denominator and numerator of $S$, Eq.~\ref{eq:s}. We assumed that nanoparticles do not change the band structure of the Si. It is known that in Si with narrow nanoparticle spacing, confinement effect leads to flattening of the conduction band~\cite{shi2015electronic, cruz1996morphological} and increases the effective mass~\cite{cruz1996morphological}, making transport coefficients different from the bulk Si. In this work, however, we consider a relatively small volume fraction of SiC particles ($\phi=0.05$) with radii make the inclusions far apart, and so the SiC nanoinclusions can be considered as perturbations encountered by the electronic wave functions of bulk Si.

%% file: Sections/model_validation_against_experiment.tex
\section{MODEL VALIDATION AGAINST EXPERIMENT}

\begin{figure*}[t]
\begin{center}
\includegraphics[width=1\textwidth]{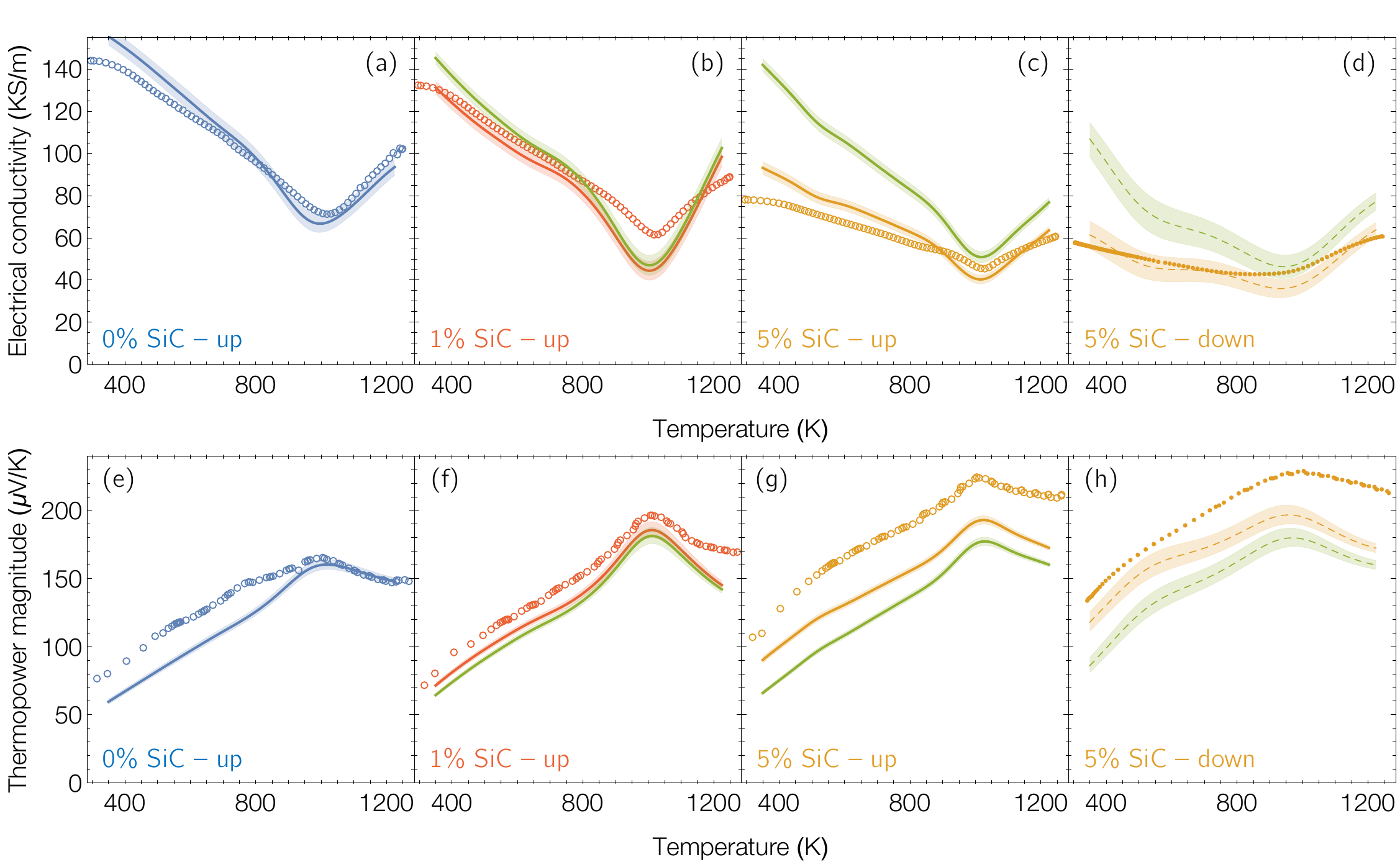}
\end{center}
\caption{Validation of the transport model by comparison with experiment. The top row (a--d) shows the experimentally measured conductivity (dots or circles), and the model prediction using the corresponding carrier concentration (solid lines with uncertainty band).  The plots (a-c) are for the heating sweep of the material with 0\%, 1\%, and 5\% SiC, and (d) is for the cooling sweep of the material in (c). The same color and symbol coding is used as in Fig.~\ref{fig:exp}. In plots (b--d) the green lines show the conductivity that would be predicted if the nanoinclusion scattering process (Eq.~\ref{eq:tau_inc}) is not included. The uncertainty bands on the prediction arise from the uncertainty in Gaussian pro cress regression of the carrier concentration. The bottom row of plots (e--h) show an equivalent comparison for the measured and predicted Seebeck coefficient.}
\label{fig:model}
\end{figure*}

Figure~\ref{fig:cc}(f) shows the electron lifetime for different scattering mechanisms in Si at 500 K with a carrier population of $2.8 \times 10^{20}\ 1/\mathrm{cm^3}$. In the parent Si, impurity scattering is dominant across the whole energy spectrum. However, scattering from inclusions has a stronger energy dependence than impurity scattering, and so the addition of inclusions creates a range of low energy states for which inclusion scattering is dominant. This fulfills the strategy for electron energy filtering illustrated in Fig.~\ref{fig:cc}(a)-(b). This is important as it shows that unequivocally the enhancement in the Seebeck coefficient in the experiments is at least partially due to the filtering effect. Moreover, the lifetimes plotted in Fig.~\ref{fig:cc}(f) indicate that the additional grain boundary scattering rate is nearly two orders of magnitude weaker than that required to explain the experimental results (Fig.~\ref{fig:exp}). The conclusion that grain boundaries are not influencing the Seebeck coefficient is also constant with other works in the literature~\cite{PhysRevB.86.115320}. Figure~\ref{fig:model} shows the comparison of the experimentally measured $\sigma$ and $S$ with the predictions from the model using the fits to the measured carrier concentrations. The predictions for the electrical conductivity are in very good agreement with the experiment. For the material with 5\% inclusions, omitting the inclusion scattering results in a significantly poorer prediction of $\sigma$. The model predictions for $S$ are less good, systematically underestimate the Seebeck coefficient by around 10\%. The Seebeck coefficient is highly sensitive to the band curvature and so we speculate that our model's underprediction of $S$ stems from the limitation in using a single band approximation to accurately represent the energy states in the whole Brillouin zone -- see for instance~\cite{minnich2009modeling}. Nonetheless, the model predicts the correct overall trends in $S$ between the different nanocomposites, and it can again be seen that the predictions are worse if inclusion scattering is omitted.

\begin{figure*}[t]
\begin{center}
\includegraphics[width=1\textwidth]{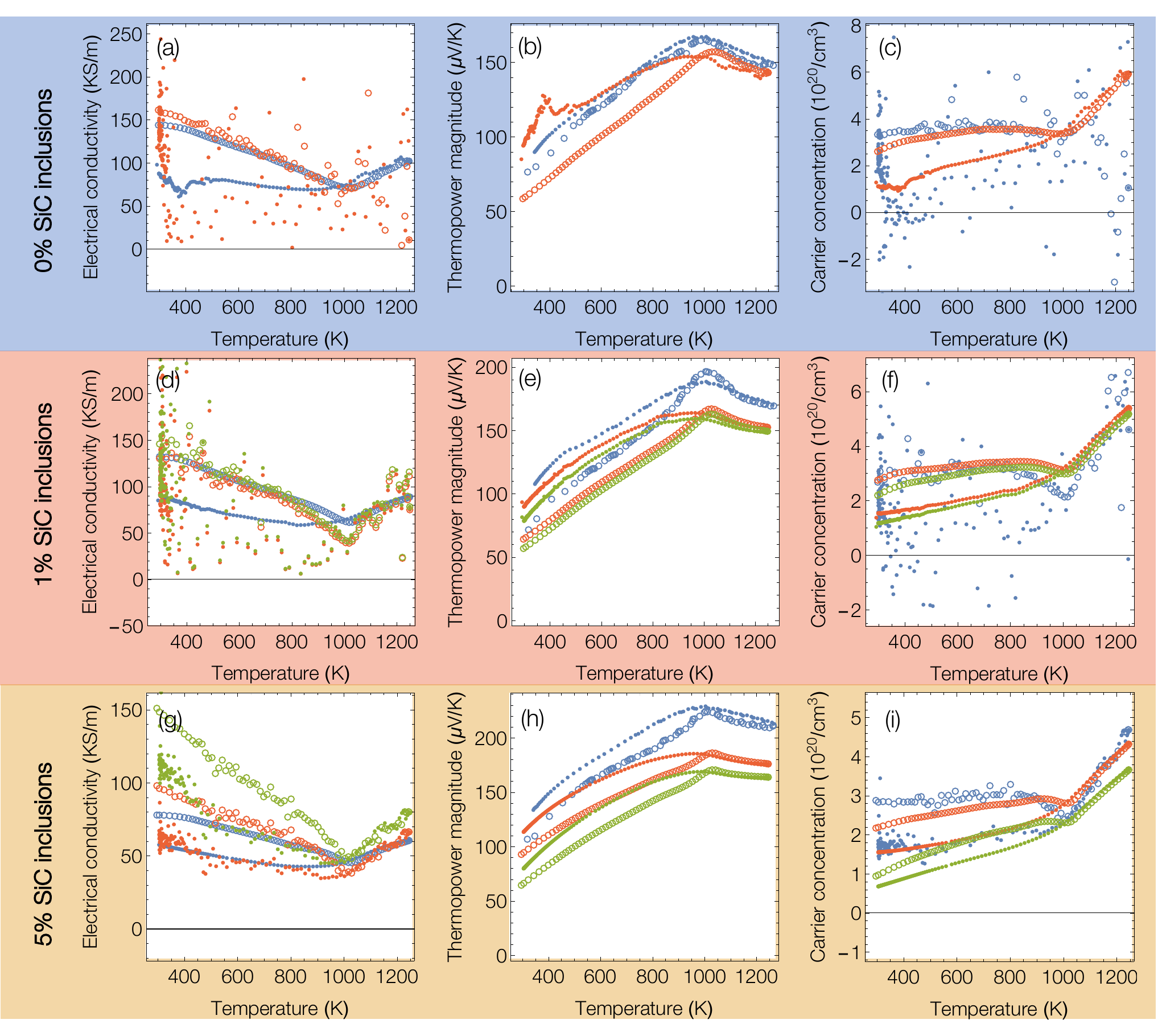}
\end{center}
\caption{Predictions of the transport model working backward from the experimentally measured electrical conductivity. The top, middle, and bottom rows show the measured data from both the heating and cooling sweeps for the material with 0\%, 1\%, and 5\% volume fraction of SiC respectively. Plots in the left column (a, d, g) show the measured electrical conductivity (blue) and the conductivity predicted by the transport model using the raw (unsmoothed) carrier concentration measurement data (red). Open and closed circles are used to distinguish the data from the heating and cooling sweeps, respectively. Plots in the central column (b, e, h) show the magnitude of the experimentally measured Seebeck coefficient (blue), and its prediction using the carrier concentration that is obtained from the transport model by back interpolation from the experimentally measured conductivity (red). The plots in the right-hand column (c, f, i) show the raw experimental carrier concentration (blue), and the carrier concentration predicted by back interpolation from the experimentally measured conductivity (red). In the second two rows, data plotted in green is the model prediction if the inclusion scattering term is omitted.}
\label{fig:validation-backwards}
\end{figure*}

Our transport model accounts for four sources of scattering and contains no tuning parameters. The agreement in Fig.~\ref{fig:model} provides strong validation the model captures most of the phenomena, quantitatively, and that there is an energy filtering effect that arises from the inclusions. However, it is also apparent that there is some evolution of the microstructure during the heating and cooling cycle that suggest the presence of other scattering processes that we have not accounted for in our model. This can be seen in Fig.~\ref{fig:validation-backwards} where instead of using the experimentally measured carrier concentration as input to the model, which is very noisy, we have used the measures conductivity as input and back interpolated the carrier concentration. This approach has the advantage that we can compare the heating and cooling data for all three materials. Looking particularly at $\sigma$ and $S$ for the parent Si with no inclusions (Figs.~\ref{fig:validation-backwards}(a) \& (b)) the experimental data shows an irreversible change in $\sigma$ from heating to cooling, that according to the transport model is consistent with the change in carrier concentration; however, in the experimental data there is not an accompanying rise in $S$ when that carrier concentration falls. This indicates that even without inclusions there is likely some source of electron filtering in the as-sintered Si matrix that is going away with annealing. the phenomenon is likely present in all samples, but with SiC inclusions we also see an electron filtering enhancement to $S$ that is not removed with heating, and we attribute this to the inclusions.

A final point to note from the plots in Figs.~\ref{fig:validation-backwards}(b), (e) \& (h) is that our model predicts $S$ of Si without inclusions self-consistently from $\sigma$ at all temperatures, with the addition of inclusions we underestimate $S$, particularly at high temperature. This suggests that the energy filtering effect obtained from inclusions is stronger than our model of the energy-dependent scattering by nanoinclusions provides.

%% file: Sections/maximum_theoritical_power_factor_enhancement.tex
\section{MAXIMUM THEORETICAL POWER FACTOR ENHANCEMENT OBTAINABLE FROM ENERGY FILTERING} \label{sec:ideal}

In the prior sections, we have validated our transport model and used it to demonstrate that SiC inclusions in Si provide an energy filtering effect. Here we extend the use of the transport model to examine how much scope exists for improving the power factor of Si nanocomposites if one were able to engineer inclusions or some other scattering mechanism that provides \emph{ideal} filtering. The approach of electron energy filtering is to shift the transport distribution’s center of mass by tuning $\tau$ -- in essence, blocking or filtering low energy electrons so that they do not contribute to transport. The concept has the practical advantage that it is easier to add extrinsic sources of scattering to the material than it is to tune the material’s band structure. However, typically one can only add sources of electrical scattering which means that, as with tuning $\chi$, enhancement of $S$, is obtained at the expense of $\sigma$. This can still yield an improvement in $ZT$ providing the power factor is increased. When tuning the parameters in $\chi$ it is found that the trade-off between $S$ and $\sigma$ leads to a narrow window of optimal conditions that maximize the power factor. However, as we demonstrate below, no such trade-off exists when tuning $\tau$. Electron energy filtering can always be made to increase $S^2$ more quickly than $\sigma$, enabling the power factor, in theory, to be increased indefinitely.

\subsection{Cutoff Model -- Perfect Filtering}

To better understand the enhancement in power factor that can be obtained with the energy selective filtering we can consider the extreme case where all the electrons with energy lower than $U_o$ are completely blocked. For this perfect filtering case, it is useful to evaluate the fraction of the total transport integrals in $\chi$ and $\gamma$ as a function of energy above the conduction band edge. We call these $X(E)$ and $\Gamma(E)$ and define them as
\begin{equation}\label{eq:X}
    X(E)=\frac{\int_0^E dE'\chi(E')\tau_p(E')}{\int_0^\infty dE'\chi(E')\tau_p(E')}
\end{equation}
\begin{equation}\label{eq:Gamma}
    \Gamma(E)=\frac{\int_0^E dE'\gamma(E')\tau_p(E')}{\int_0^\infty dE'\gamma(E')\tau_p(E')}
\end{equation}
The function $X(E)$ is always positive, while $\Gamma(E)$ is negative for $E<E_A$, where $E_A$ is the energy level at which $\int_0^{E_f} dE \gamma(E)=- \int_{E_f}^{E_A} dE \gamma(E)$. The function $\Gamma(E)$ is always less than $X(E)$ and converges to 1 more slowly than $X(E)$. The integral fractions describe how much an energy range in the conduction band contributes to transport. They can be used to define, $\alpha_\sigma (U_o)$, the ratio of a materials’ electrical conductivity with and without energy filtering. Similarly, we define $\alpha_S (U_o)$ and $\alpha_{PF} (U_o)$
\begin{equation}\label{eq:alpha_sigma}
\alpha_\sigma(U_o) = 1-X(U_o)
\end{equation}

\begin{equation}\label{eq:alpha_s}
\alpha_S(U_o) = \frac{1-\Gamma(U_o)}{1-X(U_o)}
\end{equation}

\begin{equation}\label{eq:alpha_pf}
\alpha_{PF}(U_o) = \frac{(1-\Gamma(U_o))^2}{1-X(U_o)}
\end{equation}

The power factor enhancement, $\alpha_{PF} (U_o)$, always rises above unity for small values of $U_o$, but falls back below one when $\Gamma(U_o)=1- \sqrt(1-X(U_o))$. The optimum filtering threshold for power factor enhancement, $U_{PF}^*$, satisfies the condition
\begin{equation}\label{eq:pf_star}
U_{PF}^*=E_f-\frac{eTS}{2} \left(\frac{1-\Gamma(U_{PF}^*)}{1-X(U_{PF}^*)}\right) = E_f-\frac{eTS}{2} \alpha_S(U_{PF}^*)
\end{equation}
where $S$ is the Seebeck coefficient in the bulk material. This implies that $E_f<U_{PF}^*<E_A$. Figure~\ref{fig:ideal}(a) shows the variation of power factor with carrier concentration for ideal electron filtering in Si as a function of energy filtering cutoff, $U_o$, at 500 K. This suggests that theoretically there is considerable room for further improvement to the power factor. More surprising is how the maximum obtainable power factor changes with carrier concentration as plotted in Fig.~\ref{fig:ideal}(b) -- it shows that if one can tune the energy filtering threshold then one should adopt a new paradigm for the design and optimization of thermoelectric materials. In the traditional picture of a \textit{n}-type thermoelectric, the material’s power factor is maximized at carrier concentrations that place the Fermi energy at the conduction band edge. However, if one applies perfect energy filtering one can obtain a greatly enhanced power factor by doping the material so as to push the Fermi energy deeper into the conduction band — even to doping levels where thermal conductivity is dominated by electron transport. This presents a new strategy for designing thermoelectric which is freed from the constraint of optimizing the carrier concentration.

\subsection{Generalized Cutoff Model}

\begin{figure*}[t]
\begin{center}
\includegraphics[width=0.8\textwidth]{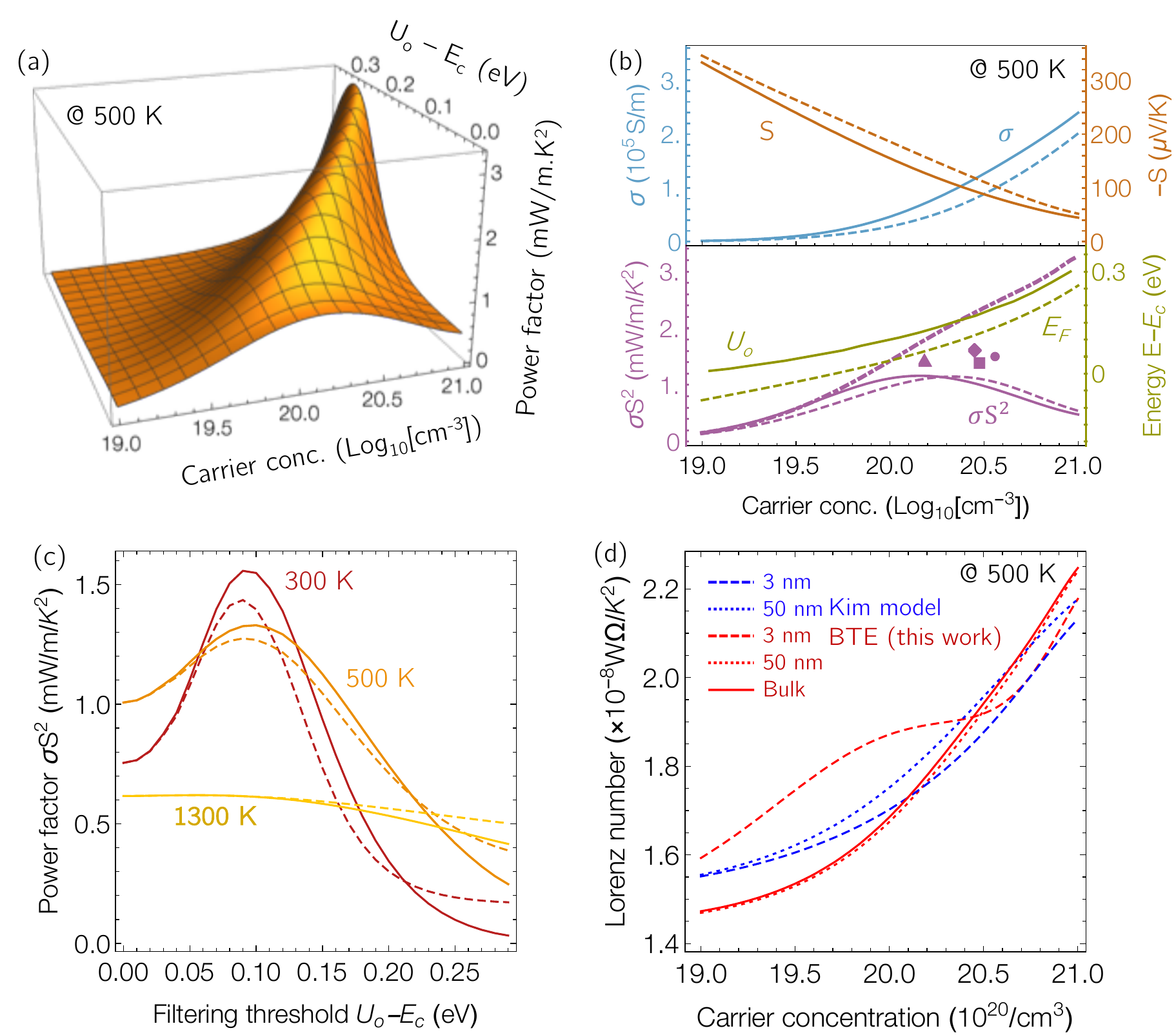}
\end{center}
\caption{Plot (a) shows the variation of power factor of P-doped Si with carrier concentration as a function of the electron perfect filtering threshold, $U_o$ at 500 K. Panel (b) shows the model prediction of the electrical transport properties in P-doped Si at 500 K. The top panel shows the conductivity $\sigma$ (blue, left ordinate axis) and thermopower $S$ (red, right ordinate axis), while the bottom panel shows the power factor (purple, left ordinate) and the optimal perfect filtering threshold $U_o$ relative to the Fermi level (green, right ordinate, solid and dashed lines respectively). The transport coefficients for the bulk and 5\% SiC additive dispersoids are plotted in solid and dash lines, respectively. The filled markers are the experimentally measured quantities for the bulk Si (triangle) and Si/SiC (1\% SiC in square, 5\% SiC during heating in diamond and during cooling in circle) membranes. The maximum PF through ideal filtering is plotted dot-dash line. Without exploiting energy filtering there is an optimal carrier concentration for the thermoelectric power, and the experimental materials are very close to it; however, by exploiting energy filtering it is possible to continue increasing the power factor by increasing the carrier concentration. The model predicts that at this temperature there is considerable scope for further enhancement of the power factor through ideal electron filtering. Plot (c) shows the comparison the perfect filtering model ($\mathrm{\tau_o = 0\ fs}$ solid lines) and a less perfect filtering model ($\mathrm{\tau_o = 10\ fs}$, dashed lines) as a function of the energy filtering threshold $U_o$. The carrier concentration is fixed at $\mathrm{10^{20}\ 1/cm^3}$, the red, orange, and yellow plots correspond to temperatures of 300, 500, and 1300 K, respectively. Shorter $\tau_o$ is more beneficial for power factor enhancement, but the optimal $U_o$ depends on temperature. Pane (d) shows the variation of Lorenz's number with carrier concentration for in a nanocomposite containing SiC nanoparticles with radii of 3 nm and 50 nm (dashed and dotted lines respectively). The blue lines show the prediction from Kim et al.'s model and red lines the predictions from this work. The solid red line shows Lorenz's number for bulk Si. The temperature is fixed at 500 K.}
\label{fig:ideal}
\end{figure*}

While the perfect cutoff model suggests an extremely large enhancement in power factor, it is not feasible to completely block low-energy electrons, here, we instead take a phenomenological approach. Regardless of the mechanism of scattering, we assume that the scattering rate from inclusions is largest for low energy electrons and weaker for high energy electrons (consistent with Fermi’s golden model of nanoparticles scattering rate). The reverse trend is true for ion scattering and so we assume that scattering of low energy electrons is dominated by scattering from inclusions but that there can exist crossover energy $U_o$ where the scattering of electrons with energy above this threshold is predominantly from dopants. The filtering threshold $U_o$ is related to the conduction band offset, $\Delta E_C$. The simplest phenomenological model of inclusion scattering is to model the additional rate of scattering as a step function so that electrons with energy $E<U_o$ are subjected to an additional scattering process with characteristic lifetime $\tau_o$, represented mathematically as 
\begin{equation}\label{eq:tau_inc}
 \tau_{c}^{-1}= \tau_o^{-1} \vartheta(U_o-E)
\end{equation}
where $\vartheta$ is the Heaviside function. Matthiessen’s rule is used to sum the rate of this extrinsic scattering term with electron-impurity and electron-phonon scattering giving the total electron scattering rate. Using this scattering function in the transport model we can examine the model’s prediction for transport properties of nanoengineered Si. Figure \ref{fig:ideal}(c) shows the predicted power factor for a material with the carrier concentration of $\mathrm{n = 10^{20}\ 1/cm^3}$, for ideal cutoff model ($\mathrm{\tau_o = 0\ fs}$) in red and generalized cutoff model ($\mathrm{\tau_o = 10\ fs}$) in blue, if we could independently control $U_o$ and $\tau_o$. This provides several important insights. The first is that the optimal filtering threshold, $U_o$, for enhancing power factor is relatively independent of $\tau_o$ and vice versa. Figure \ref{fig:ideal}(c) also shows that if the filtering threshold is not optimal electron filtering can diminish the power factor rather than enhance it. Shorter $\tau_o$ is more beneficial for power factor enhancement, but the optimal $U_o$ depends on temperature.

%% file: Sections/model_prediction_for_electron_thermal_conductivity.tex
\section{MODEL PREDICTION FOR ELECTRON THERMAL CONDUCTIVITY}

We finish our examination of the effect of nanoscale size particles on the electrical transport coefficients by briefly discussing the electronic thermal conductivity. The $\kappa_e$ is conventionally related to $\sigma$ through Wiedemann-Franz law ($\kappa_e=L\sigma T$). The term $L$ is known as Lorenz number and varies from $\mathrm{2\times \left(\nicefrac{k_B}{q} \right)^2= 1.49\times10^{-8}\ W.\Omega.K^{-2}}$ in dielectrics with low carrier population where acoustic phonon scattering is dominant, up to $\mathrm{\nicefrac{\pi^2}{3} \times \left(\nicefrac{k_B}{q}\right)^2= 2.44\times10^{-8}\ W.\Omega.K^{-2}}$ for free electrons (degenerate limit)~\cite{kim2015characterization}. In bulk Si $\kappa_e\ll \kappa_l$ the lattice conductivity, and so inaccurate prediction of $\kappa_e$ has negligible impact on the prediction of $ZT$. However, when the material is highly nanostructured so that $\kappa_l$ is suppressed to being comparable in magnitude with $\kappa_e$, then incautious determination of Lorenz number may lead to inaccurate estimation of $ZT$. Flage-Larsen et al. proposed an analytic solution for Lorenz number based on single parabolic band model as a function of carrier concentration, temperature and effective mass~\cite{flage2011lorenz}. Kim et al. proposed a thermopower-dependent expression for the Lorenz number~\cite{kim2015characterization},
\begin{equation}\label{eq:ls}
    L = 1.5+e^{\left[-\frac{|S|}{116} \right]},
\end{equation}
where $L$ is in $10^{-8}\ W\Omega K^{-2}$ and $S$ is in $\mu V/K$. This model predicts $\kappa_e$ to within 5\% for single parabolic band/ acoustic phonon scattering assumption. Substituting the BTE Eqs.~\ref{eq:sigma} and~\ref{eq:ke} in Wiedemann-Franz law leads to following ratio for Lorenz number
\begin{equation}\label{eq:l}
L=\frac{1}{(eT)^2} \left( \frac{\int \zeta(E,T) \tau(E,T) dE}{\int \chi(E,T) \tau(E,T) dE}-\frac{\int \gamma(E,T) \tau(E,T) dE}{\int \chi(E,T) \tau(E,T) dE}\right) = \frac{1}{(eT)^2}\left(\Delta_2 - \Delta_1^2 \right). 
\end{equation}
This equation is valid for noninteractive electrons with weak coupling with impurities or lattice vibration with elastic scattering. Figure \ref{fig:ideal}(d) shows the variation in Lorenz number with carrier population and inclusion size in Si/ SiC nanocomposite system at 500 K using Eqs.~\ref{eq:ls} and~\ref{eq:l}. Overall, Kim's model and the BTE model predictions are reasonably similar.

%% file: Sections/conclusion.tex
\section{CONCLUSIONS}
To summarize, we have demonstrated a new route for the synthesis of bulk thermoelectric materials with well-dispersed nanoinclusions. The novel plasma processing route for the synthesis of SiC nanoparticles decouples the control of inclusions properties from the densification and annealing of the bulk matrix. This approach is expected to enable improved control of the inclusion size distribution and number density, and it is in principle compatible with a wide range of chemistries that are free from processing constraints of solubility, nucleation, and ripening for precipitation-derived inclusions. Overall, this gives access to a rich design palate for tuning electron energy filtering. The SiC nanoinclusions demonstrated in this work scatter phonons, but also enhance power factor by electron energy filtering. A semiclassical electron transport model was developed to elucidate the electron energy filtering effect. The model was informed by the empirically measured carrier concentration and provided an excellent quantitative fit to the experimentally measured transport coefficients as a function of temperature with no tunable model parameters. The model was extended with the addition of a phenomenological model of energy selective scattering from inclusions and used to show that energy filtering must be active in the experimental system. The model was used to explore the theoretical limit of power factor suggesting that more than two-fold increase in power factor could be achieved at 500 K for \textit{n}-type Si with a carrier concentration $\mathrm{\sim 3 \times 10^{20}\ 1/cm^3}$. More remarkable, is that the theoretical maximum obtainable power factor using electron energy filtering continues to rise monotonically with increasing carrier concentration, substantially altering the constraint imposed by carrier concentration on the design of thermoelectric materials.